# A facile synthesis method and electrochemical studies of hierarchical structured MoS$_2$/C-nanocomposite

Zhenyou Li[a,b,#], Alexander Ottmann[a], Elisa Thauer[a], Christoph Neef[a], Huazheng Sai[b], Qing Sun[a], Krzysztof Cendrowski,[e] Hans-Peter Meyer[d], Yana Vaynzof[a,c], Ewa Mijowska,[e] Junhui Xiang[b,*], and Rüdiger Klingeler[a,c,°]

[a]Kirchhoff Institute of Physics, INF 227, Heidelberg University, Germany

[b]College of Materials Science and Opto-Electronic Technology, University of Chinese Academy of Sciences, Yuquan Road 19A, Beijing, 100049 China.

[c]Centre for Advanced Materials (CAM), INF 225, Heidelberg University, Germany

[d]Institute of Earth Sciences, Heidelberg University, D-69120 Heidelberg

[e]Department of Nanotechnology, Faculty of Chemical Technology and Engineering, West Pomeranian University of Technology, Piastow 45 Av., 70-311, Szczecin, Poland

*xiangjh@ucas.ac.cn

#zhenyou.li@kip.uni-heidelberg.de

°klingeler@kip.uni-heidelberg.de

A uniformly coated MoS$_2$/Carbon-nanocomposite with three-dimensional hierarchical architecture based on carbonized bacterial cellulose (CBC) nanofibers is synthesized by a facile one-step hydrothermal method followed by thermal annealing at 700 °C in Ar atmosphere. Strong hydrogen bonds between the Mo precursor and the BC nanofibers are found to be crucial for the in-situ growth of MoS$_2$ nanosheets on the nanofibers during hydrothermal process. The unique structure was maintained and the connection between MoS$_2$ and nanofibers were strengthened in the sintering process, leading to an improved stability of the resulting nanocomposite upon electrochemical cycling. The low-cost and environmentally friendly 3D web-like structure enables binder-free and carbon-free electrodes for lithium-ion batteries, which exhibit high specific discharge capacities up to 1140 mAh/g at the C-rate of 1 C without significant capacity fading for over 50 cycles. The porous conductive hierarchical structure of the composite endows excellent rate performance by avoiding the aggregation of the MoS$_2$ nanosheets and by accommodating mechanical stress which appears upon electrochemical cycling.

**Introduction**

Recent years, Molybdenum disulphide (MoS$_2$) has attracted significant attention because of its unique properties and wide field of applications, [1-6] especially the excellent electrochemical performance. Analogously to graphene, it has a two-dimensional layered structure, giving rise to unique electronic properties[5, 7, 8] and allowing for the intercalation of lithium ions without obvious volumetric change.[9-11] In LIBs, $MoS_2$ can also react with Li to form metal Mo and $Li_2S$ during a conversion process, leading to a significant improvement in capacity (670 mAh/g) as compared to intercalation based anode materials.[12] This is not restricted to LIBs as $MoS_2$ can also host Na ions which renders it a potential anode material for reversible sodium ion batteries as well.[13] However, the electrochemical performance of bulk $MoS_2$-based anodes still features disadvantages concerning practical application in LIBs. The large volume expansion accompanying the conversion reaction causes electrode pulverization and derogates the integrity of the active material and current collector.[14] Poor cycling stability and rate performance is further increased by intrinsically poor electrical and ionic conductivities of the materials, which are further deteriorated by the structure destruction during the lithiation-delithiation process.[9]

There are mainly two common approaches that have been successfully applied to overcome these issues in a variety of LIB electrode materials: downscaling of primary particles and fabrication of hierarchically structured hybrid materials.[15] In the case of $MoS_2$, various nanostructures (such as nanotubes, hollow spheres, core-shell structures etc.)[9, 12, 16] have been produced aiming at accommodating mechanical strain upon electrochemical cycling as well as at shortening the transport distance of electrons and lithium ions. In addition, several approaches for synthesizing hierarchically structured $MoS_2$/conductive material composites with nanomaterials like graphene, carbon nanotubes (CNTs) etc. have been reported.[14, 17-24] Chen *et al.*[14] reported a l-cysteine-assisted method to synthesize $MoS_2$/graphene composites with excellent cycling stability and high-rate capability. Also, cylindrical, nanostructured $MoS_2$/CNT composites showed superior electrochemical performance in LIBs.[17]

In hierarchical composites, the individual 2D $MoS_2$ nanosheets are to be preserved by the primary architectures, providing highly exposed surface area and large specific surface area for electrochemical reactions. The particular hierarchy and morphology is crucial for the electrochemical performance of the material. This also holds for preparation of the working electrode, which usually includes mixing with additional carbon and an organic binder. Even upon strong stirring it is difficult to achieve homogeneous anode slurries. Note, that strong stirring may destroy the nanostructure. Moreover, most of the applied binders are low conductivity polymers, which reduce the performance of the electrode. In this regards, it is most appealing to aim at electrodes free of binders and additional carbon. However, only very few studies on binder-free $MoS_2$ anodes including nanostructured carbon have been reported so far.[10, 25, 26]

Herein, a facile synthetic method to fabricate a composite of carbonized bacterial cellulose (BC) nanofibers uniformly coated with $MoS_2$ nanosheets (S-$MoS_2$/BC) is presented. The strong interaction between the $MoS_2$ nanosheets and the fibers is attributed to hydrogen bonds between the subsystems of the composite. This synthesis approach hence results in a stable and conductive 3D network of nanofibers, which fastens the electron and ion transport, prevents agglomeration of the nanosheets, and provides a large specific surface area. The novel architecture exhibits greatly improved cycling performance and rate performance with regard to the pure $MoS_2$ assembly. The synthesis procedure mainly applies hydrothermal synthesis using the $MoS_2$ precursors together with the BC matrix, which is a facile and easily scalable method. This method enables preparing binder-free electrodes without additional carbon, i.e. the pristine product material can be used in LIBs without further C-functionalisation. The straightforward synthesis approach applied here hence renders the material much more cost-efficient as compared to graphene and carbon nanotubes-based nanocomposites which require complex and expensive synthetic procedures.[27]

**Experimental**

**Material Synthesis**

The bacterial cellulose (BC) nanofiber matrix was prepared from the delicious food *Nata-de-coco* (Kuangquan Food Co. Ltd).[28] The *Nata-de-coco* were immersed in deionized water for 4 h to remove the sugar inside. The washed BC hydrogels were heated to 90 °C in a concentrated NaOH solution and were kept there for 6 h. Then, the hydrogels were washed for 10 h in a beaker with 100 ml water while stirring. The water was exchanged every 2 h. Afterwards, the $MoS_2$/BC composites were produced by hydrothermal synthesis. 200 mg ammonium molybdate tetrahydrate (AMT, Sigma-Aldrich, 81.0 – 83.0% $MoO_3$ basis) was dissolved in 50 ml deionized water. 200 mg thiourea (Sigma-Aldrich, 99%) was added subsequently. After that, the BC matrix was immersed into the solution under mild stirring overnight. Then, the solution as well as the BC matrix were transferred into a 50 ml stainless steel autoclave and heated in an electric oven at 200 °C for 24 h. In order to remove any remaining precursors and soluble impurities, the products were washed with deionized water for at least 12 h under magnetic stirring. The $MoS_2$/BC composites were dried through a freeze-drying method for 24 h. The final product, the S-$MoS_2$/BC composite, was prepared by sintering the $MoS_2$/BC in a tube furnace at 700 °C for 10 h in Ar atmosphere. Pure $MoS_2$ nanosheets which were studied as a reference were prepared by the same hydrothermal method without the BC matrix, followed by sintering.

**Characterization**

X-Ray powder diffraction (XRD) was performed in Bragg-Brentano geometry (Bruker-AXS D8 ADVANCE ECO) applying Cu-$K_{\alpha 1}$ radiation ($\lambda$ = 1.54056 Å). The step size $\Delta 2\vartheta$ was 0.02°. The morphology of the powder was characterized by means of a ZEISS Leo 1530 scanning electron microscope (SEM) after preliminary plasma sputtering of a 10 nm gold thin film on the samples in Ar atmosphere. X-ray photoemission spectroscopy (XPS) was carried out in an ESCALAB 250Xi ultra-high vacuum system. Using an Al $K_\alpha$ radiation source (*hv* = 1486.6 eV), a 900 µm spot size and 20 eV pass energy. The samples were prepared on a copper plate with 10 mm in diameter. Three spots were measured on each sample. The morphology of the samples was also investigated by a FEI Tecnai F30 transmission electron microscope (TEM) with a field emission gun operating at 200 kV and Energy-dispersive X-ray spectroscopy (EDX) as one mode.

**Electrochemical measurements**

Electrochemical studies were carried out using Swagelok-type cells.[29] In the case of $MoS_2$/BC and S-$MoS_2$/BC, the electrode materials were prepared by soaking the active materials in anhydrous 1-methyl-2-pyrrolidinone (NMP, Sigma-Aldrich, 99%). The slurry was pasted on a circular Cu plate (approx. 12 mm in diameter), dried overnight under vacuum at 75 °C and pressed. The resulting electrode was dried again in a vacuum oven at 75 °C for 24 h, and transferred into an Ar atmosphere glove box. The two-electrode Swagelok-type cell was assembled in the glove box using lithium foil as counter electrode and 1 M $LiPF_6$ in a 1:1 mixture of ethylene carbonate and dimethyl carbonate as the liquid electrolyte (Merck LP30). In contrast, the pure $MoS_2$ nanosheet-based electrodes were prepared from a mixture of the active material, carbon black (SuperP, Timcal) and polyvinylidene fluoride (PVDF, Sigma-Aldrich, 99%) binder with a weight ratio of 8:1:1, soaked in anhydrous NMP. The subsequent drying and pressing as well as the assembly process were the same as described above. Cyclic voltammetry and galvanostatic cycling of the cells were performed at 25 °C between 0.01 and 3.0 V versus $Li^+$/Li at various scan/current rates using a VMP3 multichannel potentiostat (Bio-Logic SAS). Electrochemical impedance spectroscopy (EIS) was carried out also by using the VMP3 multichannel potentiostat in the frequency range of 100 kHz to 0.1 Hz.

## Results and discussion

The XRD patterns of pure MoS$_2$, MoS$_2$/BC and S-MoS$_2$/BC are shown in Fig. 1. The diffraction peaks of the pure MoS$_2$ sample (blue curve) match the hexagonal phase of MoS$_2$ (space group *P6$_3$/mmc*, JCPDS card No. 37-1492).[30] Following Ref. [[31]], the detected peaks are assigned to the (002), (100) and (110) planes of MoS$_2$, respectively. For MoS$_2$/BC (red curve), the broad peak at around 20° corresponds to the BC matrix.[28, 32] The characteristic peaks of MoS$_2$ are broad and of relatively low intensity which implies poor crystallinity of MoS$_2$. In order to increase the crystallinity of MoS$_2$, the composite was sintered at 700 °C for 10 h under Ar atmosphere. The XRD pattern of the sintered material S-MoS$_2$/BC (black curve) indeed shows a sharp and pronounced (002) peak which is typical for well-stacked layers of MoS$_2$ nanostructures[31]. The peak position at 2θ = 14.25° corresponds to a *d*-spacing of 0.62 nm. As the peak width depends on the crystallinity and the strain, the (002) peak indicates a crystalline correlation length of the order of 5 nm, i.e. of about 8-10 ordered MoS$_2$ layers.

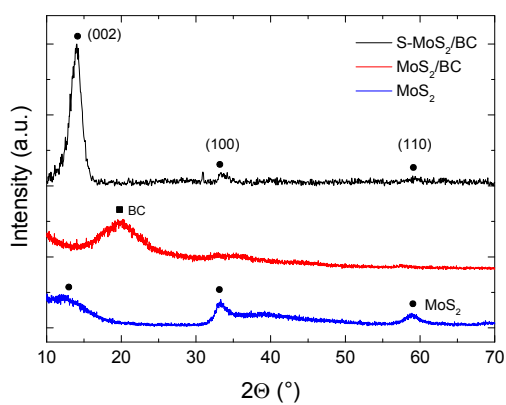

Figure 1. XRD patterns of S-MoS$_2$/BC (black), MoS$_2$/BC (red) and pure MoS$_2$ nanosheets (blue). Circles indicate Bragg peaks associated to the hexagonal MoS$_2$-phase (space group *P6$_3$/mmc*)[31] while cubic indicated the diffraction peaks of BC.

The surface chemical state and composition of the S-MoS$_2$/BC composite were investigated by XPS analysis. The XPS survey spectrum in Fig. 2a confirms the presence of C, Mo, S, and O in the sample while the Cu signal originates from the substrate. In the high-resolution Mo 3d spectrum at binding energies of 223-235 eV (Fig. 2b), the peaks at 229.2 eV and 232.3 eV correspond to the 3d$_{5/2}$ and 3d$_{3/2}$ lines of Mo$^{4+}$, respectively. The splitting is close to 3.1 eV, which is typical for Mo$^{4+}$-ions.[18, 33] No Mo$^{6+}$ signals are observed, indicating that no MoO$_x$ species are present in the composite. The high-resolution S 2p spectrum (Fig. 2c) exhibits peaks at 161.8 and 163.0 eV, which are associated with S$^{2-}$ 2p$_{3/2}$ and 2p$_{1/2}$. The intensity ratio of these two peaks is approximately 2:1 and the separation energy is 1.2 eV, which is typical for S$^{2-}$-ions. Quantitatively, the data (see Table S1) imply a S:Mo molar ratio of 2.24±0.03, The small discrepancy to the stoichiometric ratio can be attributed to defects at the nanocomposite surfaces.

The C 1s spectrum in Fig. 2d shows an overlapped behavior with the main peak at 284 eV which belongs to C-C and its right shoulder at 285.5 eV which was assigned to C=O bonds. In the O 1s spectrum in Fig. 2e, there are two splitting peaks. The one with lower binding energy attributes to the oxygen containing in the sintered nanofibers while the other belongs to the adsorption of oxygen on the sample surface. The C to O ratio in S-MoS$_2$/BC amounts to 3.35±0.14 which clearly exceeds the C:O molar ratio in cellulose of 1.2. Concerning the adsorption oxygen contribution, the real C/O ration is higher than 3.35. Therefore, we conclude that the cellulose fibers matrix is partially carbonized and reduced during the sintering process. Finally, we note that the atomic composition studied at three different spots of S-MoS$_2$/BC is quite similar, which demonstrates the uniform composition of the synthesized composite.

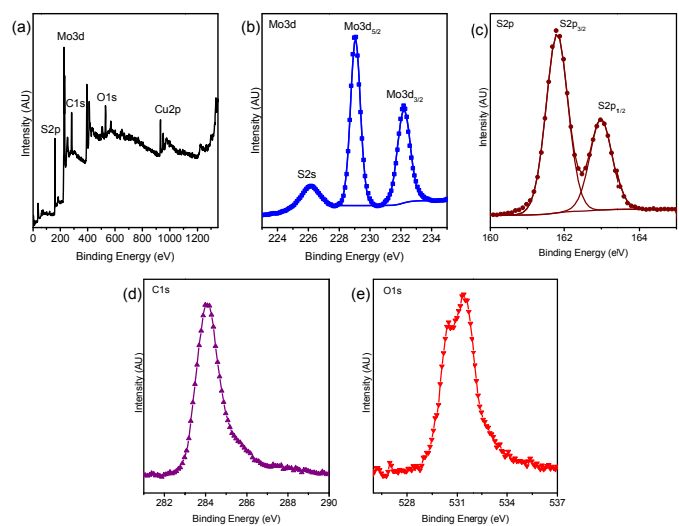

Figure 2. XPS spectra of the S-MoS$_2$/BC composite: (a) survey scan and (b) ~ (e) high resolution scans of Mo 3d, S 2p, C 1s, and O 1s respectively.

SEM images of the MoS$_2$/BC composite (Fig. 3) show that the composite displays a hierarchical structure. The pristine BC nanofibers display diameters of about 35 nm, forming a flexible, cross-linked three dimensional network (Fig. 3a). The hydrothermal synthesis step yields coating of the pristine BC nanofibers by MoS$_2$ nanosheets as illustrated in Fig. 3b-d. The thickness of the MoS$_2$ nanosheets is about 20-30 nm, which is in good agreement with the literature.[30] Applying the same hydrothermal synthesis without the BC matrix, pure MoS$_2$ nanosheets are formed which self-assemble into microspheres (Fig. S1).

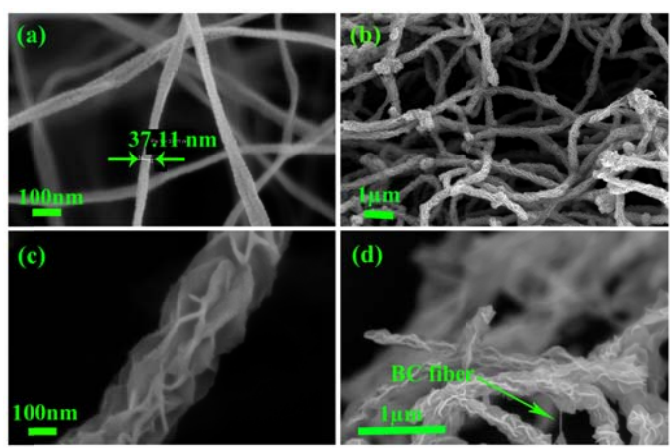

Figure 3. SEM images of pristine BC nanofibers (a) and of the MoS$_2$/BC composite (b-d).

Appropriate sintering of the material at about 700 – 900 °C is required for obtaining MoS$_2$ nanosheets with increased crystallinity. The thermal treatment should however not demolish the hierarchical structure. According to our previous study[34], the pristine BC matrix begins to decompose at around 380 °C. The decomposition temperature increases slightly if BC is combined with other materials which have a higher decomposition temperature.[32] Here, we sintered the as-prepared MoS$_2$/BC composite at 700 °C for 10 h under Ar atmosphere to carbonize the BC matrix and to boost the crystallinity of the MoS$_2$ nanosheets. The microstructure of the sintered material S-MoS$_2$/BC shown in Fig. 4 confirms that, under these conditions, the nanosheet-coated network of nanofibers is indeed preserved (cf. Fig. 3b-d).

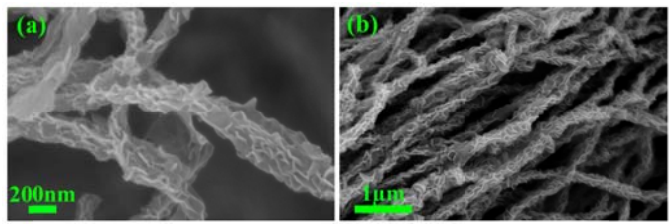

Figure 4. SEM images of S-MoS$_2$/BC composites at different magnification.

The morphology and the elemental distribution in the fibrous composites are further investigated by TEM and EDX analysis. The dispersive MoS$_2$ nanosheets on the fiber could be seen in Fig. 5a. The EDX elemental mapping in Fig. 5b clearly reveals that the elements C, Mo, and S are evenly distributed in the material in which Mo and S are restricted to the fibers. In summary, the EDX data clearly show that MoS$_2$ is present at the fibrous composite material.

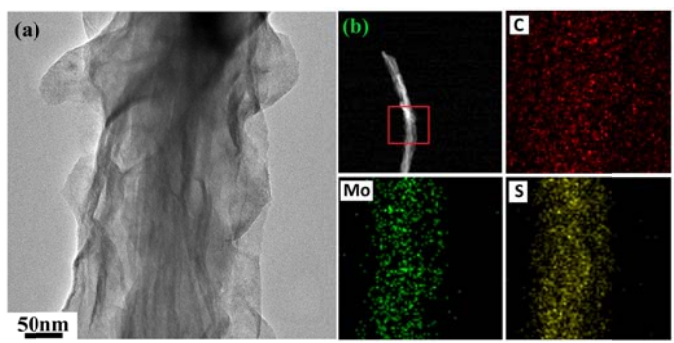

Figure 5. TEM image (a), STEM image and elemental mapping (b) of the S-MoS$_2$/BC sample.

In the following, the strong regulation effect of BC is discussed. In general, bonding of the MoS$_2$ nanosheets and the BC nanofibers is attributed to hydrogen bonds between -OH groups at the surfaces of the BC nanofibers and Mo$_7$O$_{24}^{6-}$ anions from AMT. Because of the highly active surface of the BC fibers, the surface of BC is expected to adsorb large amounts of Mo precursor.[12] MoS$_2$ nanosheets are hence supposed to grow in-situ on the surfaces of the BC nanofibers during the hydrothermal process. In order to prove these assumptions, several complementary experiments have been performed. Firstly, the Mo precursor was changed from AMT to MoS$_4^{2-}$, keeping the same hydrothermal conditions. The microstructure of the resulting material is displayed in Fig. 6b. In contrast to the AMT-based synthesis (cf. Fig. 4), the microstructure reveals self-assembled MoS$_2$-nanospheres appearing in the porous space in-between the BC nanofibers. These MoS$_2$-nanospheres are smaller than those appearing in pure MoS$_2$ obtained under the same hydrothermal conditions without BC (see Fig. S1) which is attributed to the constricted space inside the fibrous network which restrains the nanospheres' further growth to microspheres. Except for this size effect, the morphology is very similar to that of pure MoS$_2$, indicating that there is no further regulation factor of the BC matrix on MoS$_2$ nanostructures when MoS$_4^{2-}$ is used as precursor. The different effect of the BC template during the synthesis can be associated to different bonding between the BC surfaces and the precursors. Compared with Mo$_7$O$_{24}^{6-}$, MoS$_4^{2-}$ exhibits weaker hydrogen bonds with the -OH groups so that MoS$_2$ rarely growths at the BC surfaces. This result hence shows that strong hydrogen bonding is essential to modulate the growth of the MoS$_2$ nanostructures. In the case of weak BC-precursor bonding, nucleation of MoS$_2$ mostly happens in the interspace of the cross-linked BC nanofibrous network instead of at the surface of the nanofibers. Hence, the nanosheets tend to self-assemble to form spherical structures.

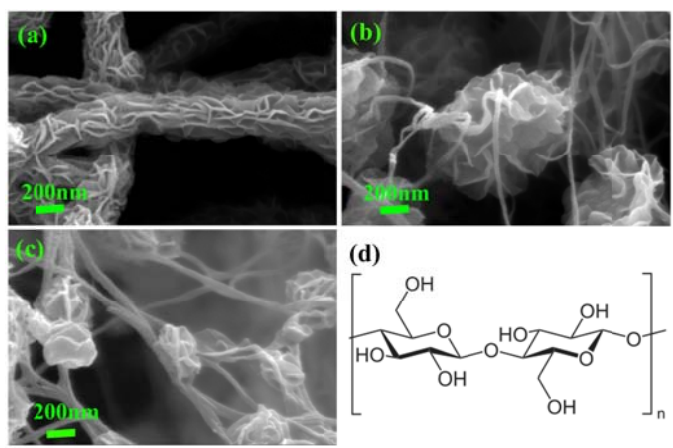

Figure 6. SEM images of MoS$_2$/BC composites produced by using different Mo precursors and matrix: (a) AMT as Mo precursor and original BC as matrix; (b) MoS$_4^{2-}$ as Mo precursor and original BC as matrix; (c) AMT as Mo precursor and H$_2$O$_2$ treated BC as matrix. (d) shows the chemical structure of BC.

To further prove the relevance of hydrogen bonding at the BC surfaces, a modified BC matrix has been used during the hydrothermal reaction. This surface modification was achieved by treating the pristine 3D BC network with a 30% H$_2$O$_2$ solution overnight at room temperature. As a result, the surface of the BC nanofibers is expected to exhibit more oxygen containing groups but lower hydrogen content.[18] Applying the original hydrothermal synthesis procedure with AMT but using the modified BC indeed yields clearly different results. After the reaction, only self-assembled MoS$_2$ aggregates within the nanofibrous network are obtained (Fig. 6c).

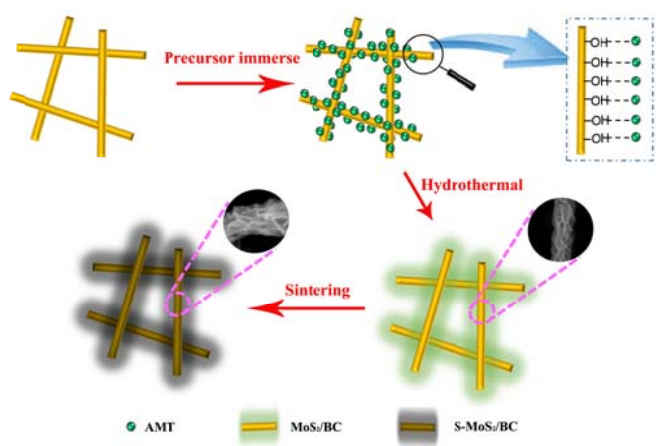

Figure 7. Schematic graph of the S-MoS$_2$/BC synthesis process.

The growth mechanism of the novel S-MoS$_2$/BC architecture is summarized in Fig. 7. The BC matrix is first immersed into the Mo precursor solution. Mo$_7$O$_{24}^{6-}$ anions from AMT form strong hydrogen bonds with the –OH groups of the BC nanofibers and disperse homogeneously on the surface of the nanofibers. This interaction further leads to nucleation of MoS$_2$ on the surface of the nanofibers, ensuring MoS$_2$ nanosheet growth on the nanofibers and preventing the agglomeration of the nanosheets into nanospheres. Abundance of hydrogen-containing surface groups (i.e. -OH groups) of the BC network is crucial for the uniform growth of MoS$_2$ nanosheets on the nanofibers. The MoS$_2$ nanosheets coated BC nanofiber structure is gradually formed upon hydrothermal treatment at appropriate temperature and pressure. Subsequent sintering of MoS$_2$/BC at high temperature under protective atmosphere yields the highly crystallized MoS$_2$ structure while the novel architecture is maintained.

To summarize, to the best of our knowledge this is the first example of a hierarchically structured nanocomposite organizing MoS$_2$ nanosheets on a flexible 3D nanofiber network. Herein, the 3D BC structure offers a porous network, leaving adequate space for the access of electrolyte for the application in LIBs. In this case, the network of carbonized BC nanofibers is supposed to act as conductive nanowires. The 3D network may also prevent the aggregation of MoS$_2$ nanosheets, endowing it with a high and stable specific surface area.

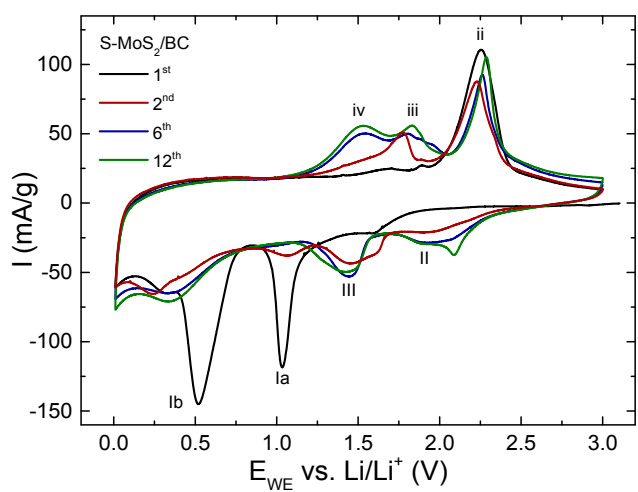

Figure 8. Selected cycles of a cyclic voltammogram of the S-MoS$_2$/BC composite obtained at a scan rate of 0.1 mV/s in the voltage range of 0.01 – 3.0 V *vs* Li/Li$^+$. Roman numbers label reduction/oxidation processes (see the text).

The electrochemical performance of the S-MoS$_2$/BC composite was studied by means of cyclic voltammetry (CV), galvanostatic cycling (GCPL), and electrochemical impedance spectroscopy (EIS). The lithium storage mechanism of MoS$_2$ is known to be strongly affected by the first lithiation half cycle, as the corresponding reduction products Mo and Li$_2$S do not react back to the initial MoS$_2$ phase. Instead, Li$_2$S is oxidized to elemental sulfur during the delithiation, and S/Li$_2$S form a redox couple during subsequent cycling. Based on ex-situ XRD at different dis-/charge states, the following reaction steps have been reported[35, 36]:

1$^{st}$ lithiation:    MoS$_2$ + $x$ Li$^+$ + $x$e$^-$ → Li$_x$MoS$_2$    (1a)

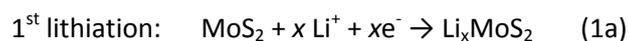

Li$_x$MoS$_2$ + (4-$x$)Li$^+$ + (4-$x$)e$^-$ → 2Li$_2$S + Mo (1b)

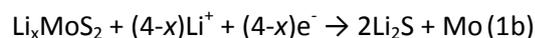

Subsequently:  Li$_2$S ↔ S + 2Li$^+$ + 2e$^-$    (2)

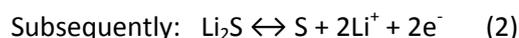

Fig. 8 shows the 1$^{st}$, 2$^{nd}$, 6$^{th}$, and 12$^{th}$ cycle of a CV, which was recorded at a scan rate of 0.1 mV/s in the voltage range of 0.01 – 3.0 V vs. Li/Li$^+$. During the initial cathodic scan, there are two distinct reduction peaks around 1.0 V (Ia) and 0.5 V (Ib), and two further features around 1.6 V and 0.3 V. The first anodic scan shows one prominent oxidation peak at 2.25 V (ii), and two minor features around 1.7 V and 1.9 V, respectively. Reduction peak Ia indicates the intercalation of Li$^+$-ions into the layered structure of $MoS_2$ (Eq. 1a), while the ensuing conversion reaction (Eq. 1b) from $Li_xMoS_2$ to Mo and $Li_2S$ is reflected in peak Ib.[35] The reductive feature around 0.3 V, which is present in all cycles, in part can be attributed to the formation of a polymeric film on the active material due to electrolyte degradation.[37-42] The prominent oxidation peak ii reflects the oxidation of $Li_2S$ to S (Eq. 2). This assignment is supported by a corresponding reductive feature (II) around 1.95 V during the 2$^{nd}$ cathodic scan. This feature evolves into a more pronounced double peak at 1.9/2.1 V upon further cycling, which is consistent with the multi-step conversion reaction of $S_8$ to $Li_2S$.[43-45] Further evidence of the above-mentioned reaction mechanism (Eq. 1-2) is the strong decrease/disappearance of the reduction peaks Ia and Ib, respectively, after the 1$^{st}$ cycle. The fact that peak Ia persists in cycle 2, albeit with far less intensity, might indicate that the initial $MoS_2$ phase was not converted completely during the 1$^{st}$ cycle. The additional redox activity, which occurs around 1.2 – 1.7 V / 1.2 – 2.0 V during the cathodic and anodic scans, respectively, has not been elucidated thoroughly in literature yet, as the involved phases appear amorphous.[36] The CV data (Fig. 8) indicate that reduction peak III at 1.45 V and oxidation peak iii around 1.8 V are related to each other, because both of them only emerge starting from the 2$^{nd}$ cycle on. Furthermore, oxidation peak iv, which evolves around 1.5 V, might be related to the reductive feature around 0.3 V, because both of them increase in intensity simultaneously from cycle 3 to 12. Possible mechanisms for these unknown redox reactions are the de-/lithiation of an amorphous $Mo/Li_2S$ matrix or amorphous $MoS_x$; $MoS_x$ (e.g. $MoS_2$) might form with excess S which could be present as the integrated intensity of the S reduction peak II is much smaller than the one of the corresponding oxidation ii.

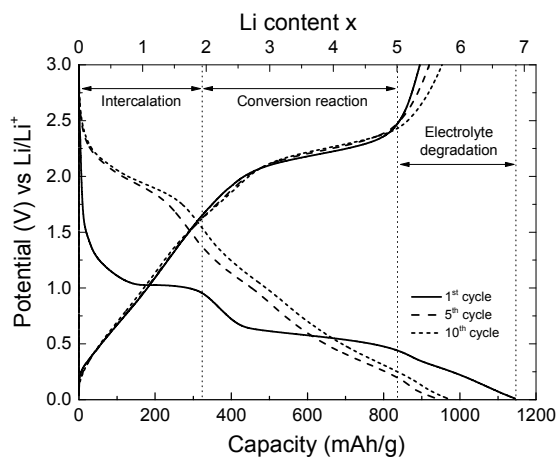

Figure 9. Electrochemical performance of the $S-MoS_2/BC$ composites: charge/discharge profiles of selected cycles at a C-rate of 1 C. The dashed lines illustrate the discharge profile by separating regions where, in the 1$^{st}$ cycle, intercalation, conversion and degradation processes dominate (see the text).

Fig. 9 shows charge/discharge profiles of $S-MoS_2/BC$ during a GCPL at a C-rate of 1 C (based on the $MoS_2$ conversion reaction, i.e. 670 mAh/g) in the potential range of 0.01 - 3.0 V *vs.* Li/Li$^+$. During the initial charge cycle, two characteristic plateaus at 1.0 V and 0.6 V are observed which is consistent with the CV data (Fig. 8), and previous reports.[46, 47] The first plateau at 1.0 V corresponds to the intercalation of Li$^+$ into the interlayer space of $MoS_2$ (Eq. 1a). It is associated to an intake of 1.9 Li$^+$/f.u.. The next plateau at 0.6 V results from the conversion reaction (Eq. 1b) with about 3 Li$^+$/f.u. intake. The composite exhibits an initial charge capacity of 1145 mAh/g at 1 C. The charge profile changes considerably in the subsequent cycles, indicating the irreversible transformation of the active material, but remains rather stable from cycle 5 on. At this stage, the plateau-like features around 2.0/2.2 V during charge and discharge, respectively, indicate the formation of Li$_2$S and its oxidation back to metallic S (Eq. 2).

The cycling performance, displayed in Fig. 10 for 50 cycles, is determined from GCPLs at 1 C between 0.01 and 3 V. S-MoS$_2$/BC shows high specific discharge capacities between 900 mAh/g and 1100 mAh/g, and an enhanced cycling stability compared to other MoS$_2$-based anode materials.[12, 46]. Both the charge and discharge capacity do not fade during the first 40 cycles and even show a slight increase. This phenomenon occurs in many other kinds of conversion reaction materials such as Fe$_3$O$_4$ and CoO due to the reversible growth of a polymer/gel-like film caused by electrolyte decomposition, which leads to increased capacities during cycling.[37, 48] Moreover, the S-MoS$_2$/BC composite exhibits a coulombic efficiency of >95% from the second cycle on. Compared with our pure MoS$_2$ and non-sintered MoS$_2$/BC samples, the S-MoS$_2$/BC hierarchical structure shows enhanced specific capacities and improved cycling stability. The charge/discharge capacities of pure MoS$_2$ begin to fade after the 1$^{st}$ cycle, decreasing to about 40% of the initial discharge capacity after 50 cycles. The cycling performance improves when BC nanofibers are used to form a MoS$_2$/BC composite. However, its capacity is only stable during the first 15 cycles and begins to decrease afterwards. Sintering at 700 °C to obtain S-MoS$_2$/BC with a carbonized nanofiber matrix further improves the cycling performance, as described above.

The observed good cycling stability is associated to a well preserved morphology upon cycling and hence to the strong bonding between BC and MoS$_2$ which is particularly enhanced by the final sintering step. This is evident from Fig. 10 a and b, which shows the microstructure of S-MoS$_2$/BC before and after 50 charge/discharge cycles. The fibrous structure is still preserved after 50 cycles. However, there are pronounced changes regarding the active coating materials as the MoS$_2$ nanosheets are altered to nanoparticles during the charge/discharge process. We conject that these changes lead to the capacity fade starting after 40 cycles.

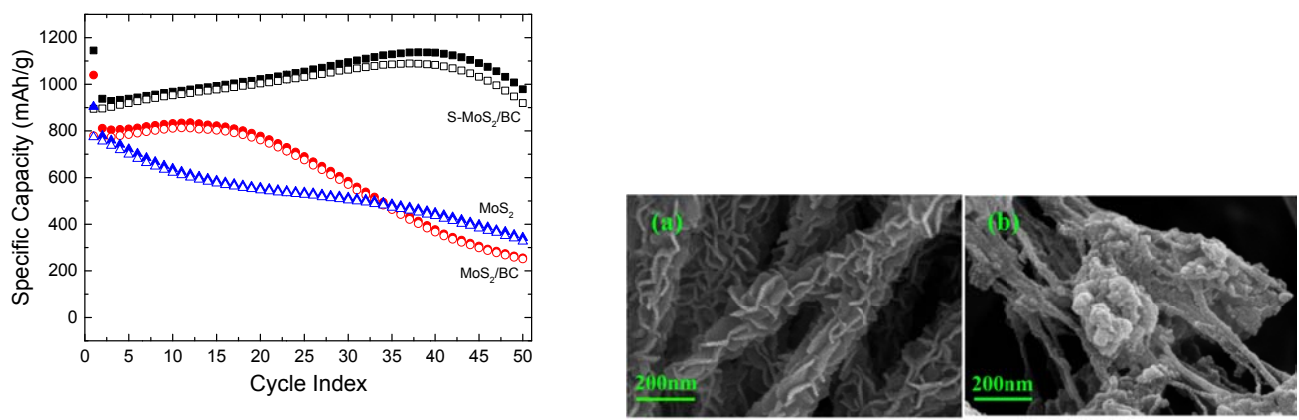

Figure 10. Top: Cycling performance of S-MoS$_2$/BC, MoS$_2$/BC and pure MoS$_2$ at the C-rate of 1 C. Bottom: SEM images of S-MoS$_2$/BC before cycling (a) and after 50 charge-discharge cycles (b).

The rate performance shown in Fig. 11 further illustrates the excellent electrochemical performance of S-MoS$_2$/BC. The discharge capacity retains 1067 mAh/g after 5 cycles at a C-rate of 0.1 C. When the current rate is increased to 0.2 C, 0.5 C, 1 C, 2 C, 5 C and 10 C, the corresponding discharge capacities are 1086 mAh/g, 995 mAh/g, 900 mAh/g, 837 mAh/g, 751 mAh/g, 671 mAh/g after 5 cycles in each case, demonstrating high capacities and cycling stability even at high current densities. Finally, when the current rate is gradually decreased back to 0.1 C, the discharge capacity recovers to 964 mAh/g which is almost the same as during the first 5 cycles at 0.1 C, even though from the 45$^{th}$ cycle on the discharge capacities suffer from the degradation effect discussed by means of Fig. 10.

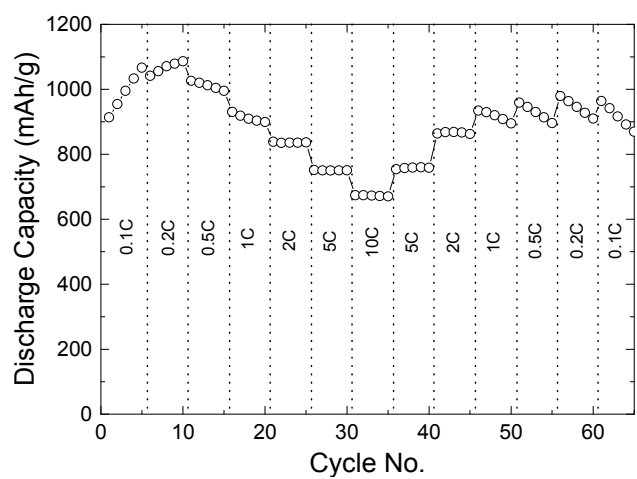

Figure 11. Rate performance of S-MoS$_2$/BC composites at different C-rates.

Electrochemical impedance spectra of pure MoS$_2$, MoS$_2$/BC and S-MoS$_2$/BC obtained at frequencies between 100 kHz and 0.1 Hz provide further insight into the electrochemical processes. The Nyquist plots of all three samples before cycling, as shown in Fig. 12a, show a depressed semi-circle at high frequencies and a straight line at low ones. The semi-circles are described by means of a generalized RC-circuit with electrolyte resistance R$_E$, charge transfer resistance R$_{CT}$, and a constant phase element CPE$_{DL}$ for the electrical double layer, using the *Z Fit* function of the *EC-Lab* software (Bio-Logic). The used equivalent circuit is shown in Fig. 12a, and the calculated parameters are listed in Table S2. The diameter of the semi-circle, which represents R$_{CT}$, is similar for both the sintered and the non-sintered BC composites, while the pure MoS$_2$ sample shows a significantly higher value. This indicates that the organic nanofiber network plays a crucial role, remarkably promoting the electron transfer in the composite electrode materials. The reduced resistance value can be attributed to direct and efficient electrical pathways between the active material and the current collector, constructed by the conductive carbon nanofiber network. The results hence confirm that the electronic kinetics is increased dramatically in the BC-based materials, which straightforwardly explains the significant enhancement of the cycling and rate performances in these materials. Comparatively, the R$_{CT}$ value of all the three samples after cycling (Fig. 12b) are decreased dramatically. And S-MoS$_2$/BC has a lower resistance than MoS$_2$/BC, exhibiting a more stable structure and better conductivity which explains the further enhanced cycling performance.[49, 50]

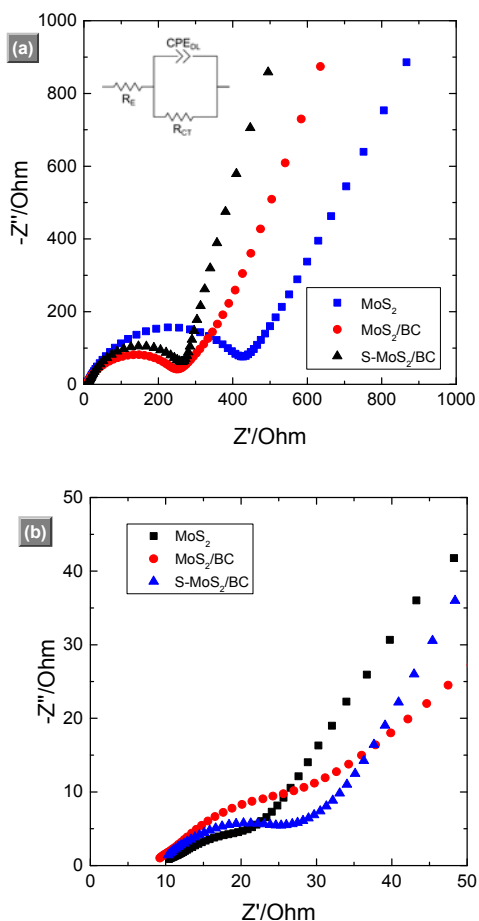

Figure 12. Nyquist plots of MoS$_2$, MoS$_2$/BC and S-MoS$_2$/BC in the frequency range from 100 kHz to 0.1 Hz: (a) before cycling, (b) after 1 cycle. The insert of (a) shows the simulated circuit.

In order to highlight the good electrochemical performance of the S-MoS$_2$/BC nanocomposite presented at hand, Table S3 lists the discharge capacities of various MoS$_2$-based anode materials after 50 cycles. Based on those values, the S-MoS$_2$/BC composite exhibits superior properties compared to other MoS$_2$/CNF composites[11, 51] and MoS$_2$-based binder-free electrodes[52, 53]. However, MoS$_2$/CNT[10] and MoS$_2$/Graphene[54] composite materials usually show higher capacities and better cycling stability than the S-MoS$_2$/BC counterpart. In contrast to those materials, the S-MoS$_2$/BC composite benefits from the simple and well-controllable synthesis process, which includes inexpensive starting materials and a product which can be used as electrode material without further additives (carbon black / binder).

**Conclusions**

In summary, we report the facile and cost-efficient synthesis of a hierarchically structured MoS$_2$/C-nanocomposite in which MoS$_2$-nanosheets are uniformly covering a flexible 3D nanofiber network of carbonized bacterial cellulose. The nanocomposite is used as a binder-free and carbon-free anode material of LIBs. It exhibits enhanced lithium storage properties including a high initial capacity (1137 mAh/g at 1 C), good cycling stability (967 mAh/g after 50 cycles at 1 C) and excellent rate performance. Such outstanding performance can be ascribed to the advantages endowed by the porous 3D network structure: tight connection between MoS$_2$ and nanofibers, and a highly conductive matrix. The strong connections between the active material and the matrix, which was derived from the hydrogen bonding of the precursors, help to buffer the volume changes upon electrochemical cycling and increase the electron kinetics, playing a key role herein. The strategy to employ low-cost, environmentally friendly, flexible, 3D web-like structure offers an effective and versatile way to design novel binder-free carbon-free electrode materials. In addition to the excellent performance in Lithium ion batteries, the S-$MoS_2$/BC with unique structure is also a good candidate for other applications demanding stable, nanoporous, and conductive $MoS_2$/C-composites such as hydrogen evolution and photocatalysis.

**Acknowledgements**

The authors thank I. Glass for technical support. Z.L. acknowledges financial support by the Chinese Scholarship Council and by the Götze foundation. The authors are grateful for financial support of the CleanTech-Initiative of the Baden-Württemberg-Stiftung (Project CT3 Nanostorage). A.O. and C.N. acknowledge support by the IMPRS-QD and the Heidelberg Graduate School of Fundamental Physics (HGSFP), respectively.